\documentclass[sigconf, nonacm, pdfa]{acmart}

\newcommand\vldbavailabilityurl{https://github.com/sdbs-uni-p/vldb25-dbms-live-patching}

\newcommand\vldbpagestyle{plain} 

\usepackage[a-2b]{pdfx}   % for PDF/A-1b

\usepackage{xargs}
\usepackage{hyperref}
\usepackage{multirow}
\usepackage{graphicx}

\usepackage{paralist} % for compact lists
\usepackage{csquotes} % always getting the quotes right

\usepackage{url}

\usepackage{siunitx}
\usepackage[inkscapearea=page]{svg}

\usepackage{pifont}

\definecolor{diffstart}{rgb}{.5,.5,.5} % Grey
\definecolor{diffincl}{rgb}{0,.5,0} % Green
\definecolor{diffrem}{rgb}{1,.27,0} % OrangeRed

\usepackage{listings}
  \lstdefinelanguage{diff}{
    basicstyle=\ttfamily\small,
    morecomment=[f][\color{diffstart}]{@@},
    morecomment=[f][\color{diffincl}]{+\ },
    morecomment=[f][\color{diffrem}]{-\ },
  }

\newcommand*{\wfpatch}{\textsc{WfPatch}}

\newcommand*{\kpatch}{Kpatch}
\newcommand*{\ksplice}{Ksplice}
\newcommand*{\kgraft}{kGraft}

\newcommand{\xdownarrow}[1]{%
  {\left\downarrow\vbox to #1{}\right.\kern-\nulldelimiterspace}
}

\usepackage{tikz}
\newcommand*\circled[1]{\tikz[baseline=(char.base)]{
            \node[shape=circle,draw,inner sep=0.4pt,fill=black,text=white] (char) {#1};}}

\usepackage[]{todonotes}

\usepackage[normalem]{ulem}

 \newcommandx{\improvement}[1]{\todo[inline,backgroundcolor=red!25,bordercolor=red]{#1}}

\usepackage[nameinlink]{cleveref}

\makeatletter
\edef\textFontName{\fontname\csname\f@encoding/\f@family/\f@series/\f@shape/\f@size\endcsname}

\makeatother

\newcommand{\includeExperimentFigure}[1]{%
            \includegraphics[width=\linewidth]{#1.pdf.trim.pdf}%
}

\usepackage{xcolor}

\begin{document}

\title{The Case for DBMS Live Patching [Extended Version]}

\author{Michael Fruth}
\orcid{0000-0003-2933-5093}
\affiliation{%
  \institution{University of Passau}
  \streetaddress{Innstraße 43}
  \city{Passau}
  \state{Bavaria}
  \postcode{94032}
  \country{Germany}
}
\email{michael.fruth@uni-passau.de}

\author{Stefanie Scherzinger}
\orcid{0000-0002-1960-6171}
\affiliation{%
  \institution{University of Passau}
  \streetaddress{Innstraße 43}
  \city{Passau}
  \state{Bavaria}
  \postcode{94032}
  \country{Germany}
}
\email{stefanie.scherzinger@uni-passau.de}

\begin{abstract}
Traditionally, when the code of a database management system (DBMS) needs to be updated, the system is restarted and database clients suffer downtime, or the provider instantiates hot-standby instances and rolls over the workload. We investigate a third option, live patching of the DBMS binary. For certain code changes, live patching allows to modify the application code in memory, without restart. The memory state and all client connections can be maintained. Although live patching has been explored in the operating systems research community, it remains a blind spot in DBMS research. In this \emph{Experiment, Analysis \& Benchmark} article, we systematically explore this field from the DBMS perspective. We discuss what distinguishes database management systems from generic multi-threaded applications when it comes to live patching. We then propose domain-specific strategies for injecting quiescence points into the DBMS source code, so that threads can safely migrate to the patched process version. We experimentally investigate the interplay between the query workload and different quiescence methods, monitoring both transaction throughput and tail latencies. We show that live patching can be a viable option for updating database management systems, since database providers can make informed decisions w.r.t.\ the latency overhead on the client side.
\end{abstract}

\maketitle

\pagestyle{\vldbpagestyle}
\ifdefempty{\vldbavailabilityurl}{}{
\vspace{.3cm}
\begingroup\small\noindent\raggedright\textbf{Artifact Availability:}\\
The source code, data, and a reproduction package
have been made 
available at \url{\vldbavailabilityurl}.
\endgroup
}
\section{Introduction}
Database management systems (DBMS) are part of the critical IT infrastructure 
and must be maintained with care. When it comes to security patches, database restarts are experienced as highly disruptive by clients, especially when long-standing  connections must be severed. At the very least, they can be untimely and force database clients to work around downtimes.

Considerable effort has been made to accelerate the restart of DBMS servers~\cite{DBLP:conf/osdi/ZhengTKL14, DBLP:conf/sigmod/GoelCGMMHW14, DBLP:conf/sigmod/CaoSSYDGW11}.  Facebook, for instance, relies on \emph{shared memory} to accelerate the restart of certain distributed systems~\cite{DBLP:conf/nsdi/NishtalaFGKLLMPPSSTV13, DBLP:conf/usenix/BronsonACCDDFGKLMPPSV13, DBLP:journals/pvldb/AbrahamABBCGMMRSWZ13}: Among these systems is Scuba~\cite{DBLP:journals/pvldb/AbrahamABBCGMMRSWZ13}, a main memory database backing about \SI{120}{\giga\byte}, for which the restart time was reduced from 2--3~hours to 2--3~minutes~\cite{DBLP:conf/sigmod/GoelCGMMHW14}. While this may be acceptable for a full upgrade, it may not be justifiable for a small (security) fix.
Alternatively, additional DBMS instances are run in parallel (e.g.\ as hot-standby or multi-master)~\cite{PostgreSQLStandby, DBLP:journals/vldb/MinhasRCASW13, MariaDBReplication, MySQLgaleraCluster, MariaDBGaleraCluster, MySQLCluster, PostgreSQLBDR}, and updates are realized by \emph{rolling over}~\cite{PostgreSQLBDR, PostgreSQLStandbyRolling, MicrosoftSQLServerRolling, AmazonUpgrade, DBLP:conf/sigmod/DagevilleCZAABC16, GaleraClusterRolling, MariaDBRolling} on these instances.

\noindent\textbf{Motivation.}
We investigate a \emph{third way}, made possible by recent advances in live patching user-space applications for Linux.
Live patching performs a code change (i.e., a \emph{patch}) directly in memory, while the software is running. Instead of a restart, the threads are gracefully patched so that they read from an updated code segment.

\begin{figure}
\centering
\includeExperimentFigure{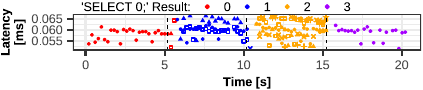}
  \caption[]{Seamlessly live patching MariaDB through four code versions. Colors distinguish the code versions.\footnotemark}
  \label{fig:patchOneByOne}
\end{figure}

\footnotetext{Throughput per worker fixed at 5 queries per second. Extreme latencies below $10^\text{th}$ / above $90^\text{th}$ percentile filtered out to improve the readability of the chart. Patching using local quiescence for the one-thread-per-connection policy, concepts to be introduced.}

We illustrate the potential of this approach for DBMSs with a micro-experiment using the \wfpatch{} framework~\cite{DBLP:conf/osdi/RommelDFKBMSL20}. We run a version of MariaDB with \emph{quiescence points} injected into the application source code. When the control flow of a thread reaches a quiescence point and a patch exists, the thread migrates to the patched version.
Figure~\ref{fig:patchOneByOne} shows the
query latencies, while the number of connections is scaled up and down. Each data point is one measurement. Every five seconds, we initiate live patching to migrate the DBMS binary to a new version. The colors indicate the individual versions, and the shapes distinguish the database connections (allowing to discern new connections). 
As can be seen by the change in color, MariaDB gracefully migrates through four code versions while maintaining all existing connections, evaluating queries, and even accepting new connections.

\noindent\textbf{State of the Art.}
In the maintenance of operating systems, live patching is established practice: IBM AIX~\cite{IBMAIX}, Windows virtual machines for Azure~\cite{WindowsVMPatch}, or different live patching tools for Linux (e.g.\ \kpatch{}~\cite{kpatch} from RedHat, \kgraft{}~\cite{kgraft} from SUSE or \ksplice{}~\cite{DBLP:conf/eurosys/ArnoldK09} from Oracle) are routinely used in production environments, showing that live patching is feasible,  even for highly critical infrastructure.

While live patching in the kernel space is state-of-the-art, 
live patching applications in user-space is still underdeveloped. Any tools publicly available are limited to research prototypes.  
In proof-of-concept evaluations, a variety of applications are considered. Although these experiments often include database management systems~\cite{DBLP:conf/usenix/MakrisB09, DBLP:conf/osdi/RommelDFKBMSL20, DBLP:conf/oopsla/HaydenSDHF12, DBLP:conf/middleware/WeichbrodtHAAK21}, the DBMS software is merely treated as yet another generic multithreaded application, the same as caches or web servers. Consequently, the special challenges of database management systems are not adequately taken into account.

Notably, several commercial database providers advertise live patching capabilities. For example, Azure allows live patching of the SQL Server Engine. A blog claims that more than 80\% of typical SQL bug fixes can be applied by live patching~\cite{AzureLivePatching}.
This indicates the vast potential for practical impact of database live patching. However, since these tools are exclusively operated in-house, they are not available to the database research community.

So far, there has been no systematic exploration of the potential of live patching for multi-threaded DBMS software from the unique perspective of database systems research.

\noindent\textbf{Contributions.}
In this paper, we systematically explore the feasibility of live patching for database management systems.

\begin{compactitem}
  \item DBMSs have unique characteristics that set them apart from other multi-threaded applications. We show that there are specific desiderata for live patching to be practical. We focus on how to statically prepare the code of database connection management; 
  we study two common policies, one-thread-per-connection and thread pools.

  \item We propose a novel approach for achieving safe thread quiescence (the prerequisite for live patching) in database thread pools. Our solution defines the order in which threads enter into quiescence based on the role of the thread (e.g., listener or worker thread). We empirically show that our approach operates without deadlocks. 

  \item For our extensive experiments, we prepared two open source DBMSs for live patching (MariaDB and Redis). We successfully apply real-world patches from GitHub and explore two alternative quiescence methods. We experimentally evaluate live patching from different stakeholder perspectives and identify the key factors determining the performance. In particular, we study the impact on tail latencies under different query workloads. Our insights enable database providers to make an informed decision w.r.t.\ live patching.  
\end{compactitem}

\section{DBMS-Specifics}
\label{sec:challenges}

\noindent\textbf{Desiderata.}
In patching a multi-threaded application, we prefer a short  \emph{synchronization time}, i.e.\ the time it takes from patch triggering until \emph{all} threads run in the patched version. Ultimately, this determines how quickly a security vulnerability can be closed.

However, database management systems constitute a family of applications with highly specific requirements that set them apart from other multi-threaded applications. We formulate these additional desiderata for patching a DBMS software binary:
\begin{compactenum}
\item The DBMS server maintains the existing client connections, and even allows for new connections to be made. 
\item Patching does not cause database deadlocks. 
\item The database state (which can be large) remains available.
\item Code patches can be applied for different query workloads.

\end{compactenum}

Not all of these desiderata can be met with traditional update methods:
In a classic system restart, connections must be severed, transactions aborted, and the database state must be restored upon restart
(costing minutes or even hours~\cite{DBLP:conf/osdi/ZhengTKL14, DBLP:conf/sigmod/GoelCGMMHW14}).
When running instances in parallel, existing client connections must be carefully handled, and a short failover time may still be noticeable.
Moreover, the hardware requirements multiply (temporarily).

In the following, we discuss the above desiderata in light of live patching and point out the technical challenges.

\noindent\textbf{Technical Challenges.}
(1)~Database connection handling is  highly system-specific. Implementations range from simplistic to complex: The key-value store Redis uses a single-threaded event loop, where all commands are executed in sequential order (since version~6, Redis supports multiple threads for I/O operations). In contrast, PostgreSQL maps each connection to its own process and forgoes multi-threading. MariaDB is multi-threaded and supports different connection management policies.
Changing the DBMS source code by injecting quiescence points is therefore best done by developers with domain knowledge.

(2) Implementing a transaction system is a delicate task. Any manipulations of the source code that cause threads to block (as quiescence points will), amplify the risk of deadlocks. Again, developers must be highly prudent.

(3) Compared to other families of applications, database management systems can hold very large states in memory. These states are expensive to recover at system restart and may also be expensive to copy during live patching, where we need to prepare the new address space containing the code changes.

(4) Different query workloads bring about different challenges. In particular, workloads containing long-running queries are likely to be unsuitable for live patching methods in which all threads must block until they reach a global barrier. Such a \enquote{stop the world} event may even noticeably increase tail latencies.

\noindent\textbf{Stakeholders.}
In addressing these challenges, we
assume different perspectives:
From the \emph{perspective of the developer} of the DBMS, quiescence points must be injected into the source code in a safe manner. Specifically, changes in the system must not introduce new deadlocks. Moreover, changes to connection management must not noticeably deteriorate the achievable database throughput.
Note that these extensions to the DBMS code are a one-time effort (although, of course, these changes must be maintained over time). 

From the \emph{perspective of the database clients}, performance should not degrade noticeably. Ideally, database clients remain unaware that patches are being applied. In distributed settings, (tail) latencies are a particular concern, since latencies exceeding the $99^\text{th}$ latency percentile can degrade the client experience (and when they build up, even the entire system performance~\cite{DBLP:journals/cacm/DeanB13}).

We also assume the \emph{perspective of the database provider}, who has to choose between performing a restart of the application, rolling over on standby instances, or live patching. 
This requires that decision makers be able to predict the latency overhead. Consequently, we explore the key factors that determine the patch application time, such as the size of the database state and the size of the patch. 

\noindent\textbf{Scope of this work.}
In general, DBMS developers will scrutinize code changes (such as for security fixes) w.r.t.\ their suitability for live patching. As we outline in Section~\ref{sec:categorize}, there are certain technical constraints. However, developers also need to carefully assess any potential \emph{behavioral} changes. %They may then recommend suitable patches. 
Ultimately, it is the responsibility of the database provider to decide whether to apply a given patch to a database running in the production environment.

In this article, we focus on the technical aspects of implementing DBMS live patching, but not the question of whether the behavioral changes of a code change make it suitable for live patching. This warrants separate research and is beyond the scope of this article.

\section{Background -- Live Patching}
\label{sec:background}

There are several (prototypical) tools
for live patching user-space applications and we refer to our discussion of related work for an overview (\Cref{sec:relatedWork}).
In our introduction to the core concepts, we focus on the framework \wfpatch{}~\cite{DBLP:conf/osdi/RommelDFKBMSL20} by Rommel et al. 

\subsection{Quiescence Points}

Live patching changes the memory state of a process. For this to be safe, a thread must be in a well-defined state. \wfpatch{} relies on software developers to identify such safe states and impose barriers, the \emph{quiescence points}, in the application source code. Once the control flow of a thread reaches a quiescence point, the thread is either blocked, patched, or continued. The action depends on the chosen quiescence method, and we discuss two methods below.\footnote{Rommel et al.\ further propose group quiescence, which we do not consider here.} 

\subsection{Quiescence Methods}
\label{sec:backgroundQuiescenceMethods}

\noindent\textbf{Global.} A thread blocks when it reaches a quiescence point.  
Once all threads have reached their quiescence point, the patch is applied, and all threads collectively
migrate to the patched version.

\begin{figure}[t]
\centering
\includegraphics[width=\linewidth]{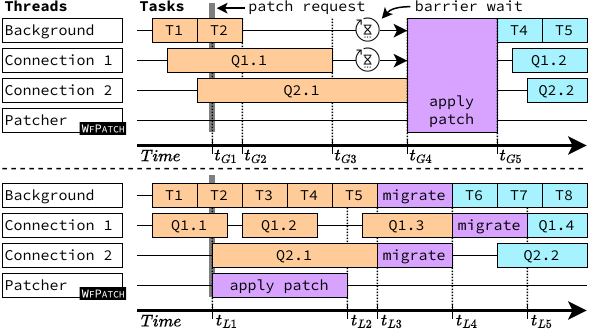}
  \caption{Live patching a DBMS: global quiescence (top) vs.\ local quiescence (bottom). Graphic based on Figure~2 of \cite{DBLP:conf/osdi/RommelDFKBMSL20}.}
  \label{fig:globalVsLocalQuiescence}
\end{figure}

The upper half of \Cref{fig:globalVsLocalQuiescence} visualizes a scenario with global quiescence. We assume a DBMS with one background thread, two threads each serving a connection, and one patcher thread. The patcher thread
is spawned by \wfpatch{} and performs all heavy-weight tasks for patch application (see \Cref{sec:wfpatchASGeneration}). 
At time~$t_{G1}$, a patch request is made. Upon completing task~$T2$, the background thread blocks. Then the thread serving connection~1 blocks. At time~$t_{G4}$, the thread serving connection~2 reaches its barrier, achieving global quiescence. The patcher thread applies the patch.
By time~$t_{G5}$,
 all threads resume work in the patched 
 version.

\emph{Drawbacks.}
In global quiescence, blocking threads can cause various problems: (1)~Long wait times: All threads reach their quiescence point timely except the thread serving connection~2. If it executes OLAP-style queries, long wait times occur.
(2)~Unbound wait times: Similar to problem~1, but now connection~2 is idle. It waits for user input to reach its barrier, causing unbounded wait times. 
(3) Deadlock: Connection~1 blocks at its barrier while holding a lock. Connection~2 is waiting for the release of this lock, resulting in a cyclic dependency and, therefore, a deadlock. 

\noindent\textbf{Local.} In this method, each thread can individually migrate to the patched process version, upon reaching its quiescence point. No blocking or waiting for other threads is needed.

Local quiescence is visualized in the lower half of \Cref{fig:globalVsLocalQuiescence}: At time~$t_{L1}$, a patch request is made, and the patcher thread prepares the patched process version. At time $t_{L2}$, the patched process version is ready and threads can migrate to this version. The thread serving connection~2 and the background thread both reach a quiescence point at time $t_{L3}$ and the migration is performed. Finally, the thread serving connection~1 reaches its barrier at time $t_{L4}$.
By time $t_{L5}$,
all threads run in the patched process version.

\subsection{Categorizing Patches}
\label{sec:categorize}
From a technical point of view, each change in source code affects a different region of the memory layout of a program. Not every patch can be applied with every live patching framework, as these are often restricted to patches that affect certain memory regions (more details in \Cref{sec:wfpatchLimitations}).

Furthermore, the semantics of a patch, i.e.\ the effect or changed behavior of the application after patching, must be well understood. While fewer problems arise due to the non-blocking property of local quiescence, this method is not universally applicable and restricted to a certain category of semantic changes. 
In discussing this next, we categorize patches according to their effect.

A \emph{thread-local patch} is a code change that affects only the given thread itself and no other threads. For example, consider a security fix that adds a boundary or a NULL pointer check~\cite{kernelOrgLivePatch}.
Thread-local patches can be applied with any quiescence method, including local quiescence.
In the case of a \emph{thread-group patch}, the changed or patched behavior affects some threads, but not all. For example, consider that the data to be processed changes in a producer-consumer scenario~\cite{kernelOrgLivePatch}, so a joint migration is essential. Therefore, we need to resort to global quiescence.
Finally, a \emph{process patch} enforces that all threads of the process are patched at the same time. This is also only feasible under global quiescence. 

Statically determining the correctness of dynamic updates is an undecidable problem~\cite{DBLP:journals/tse/GuptaJB96}. Only a skilled developer can judge the effects of a given patch and categorize it accordingly.

\subsection{\wfpatch{} Framework}
\label{sec:wfpatch}

The \wfpatch{} framework consists of a modified Linux kernel, a user-space library, and a customized version of \kpatch{}~\cite{kpatch} for patch generation. We explain the details necessary to understand this article and refer to the original article for further details~\cite{DBLP:conf/osdi/RommelDFKBMSL20}. 

\subsubsection{\textbf{\kpatch{}}}
\kpatch{}~\cite{kpatch} is a suite of tools for live patching the Linux kernel. Rommel et al.~\cite{DBLP:conf/osdi/RommelDFKBMSL20} customized the \kpatch{} tool  
to also support patch generation for user-space applications. The modified version of \kpatch{} is used for patch generation, while loading and applying the patch is handled by the \wfpatch{} user-space library.

\subsubsection{\textbf{Address Space Generation}}
\label{sec:wfpatchASGeneration}
With global quiescence, a patch is applied \emph{directly} to the address space of the process.
However, with local quiescence, multiple address space generations are managed. We outline these concepts in the following. 

Linux is divided into user-space and kernel-space: All applications of the Linux kernel run in kernel-space and application software etc.\ run in user-space (e.g.\ a DBMS). The memory, also called \emph{address space (AS)}, of a user-space application is shared between its threads. For example, the stack, heap, or .text segment (executable instructions) reside in the address space. On a low-level basis, the address space consists of a number of regions or \emph{virtual memory areas (VMAs)}. A VMA forms a contiguous memory area. Each VMA is further divided into several pages, representing the smallest unit (the typical size is \SI{4096}{\byte}). All access to memory is performed on virtual addresses which are translated based on page tables 
to physical addresses. Each process contains information about its address space, i.e.\ a list of VMAs, the page table, etc. All of this is kept in the \emph{memory map (MM)}. Thus, an address space is the abstract concept represented by the MM structure in Linux.

Linux has a strict one-to-one relationship between the memory map and the process. \wfpatch{} relaxes this so that threads of the same process can have different memory maps, i.e.\ threads can operate in different address spaces, yet the individual memory maps remain siblings.
When creating a new memory map, the memory map data structure with all its attributes of the calling thread is copied and kept in sync with its siblings by sharing pages. Logically, they are two separate address spaces, but all entries refer to the same pages. All memory maps are kept in sync  using the very first memory map as a synchronization point.
Each distinct memory map forms an AS generation, and the synchronization between AS generations can be suspended on the granularity of the VMAs. The \emph{copy-on-write (COW)} mechanism is used on VMAs that are no longer shared. Using COW, changes are no longer reflected in the other AS generations, as the page is copied when modified.

The way \wfpatch{} clones a memory map for AS cloning is similar to the fork() system call: A new process is created by duplicating the AS of the calling process. All pages of both processes are shared as read-only and marked as COW. However, there is a difference from AS cloning: All pages are shared by default (shared mapping), and changes are synchronized with all other AS generations. Only certain VMAs of the AS are marked as COW.

Once an AS has been created and the corresponding regions have been marked as read-only, a patch can be applied to create a patched AS generation: \wfpatch{} loads the patch binary file generated by \kpatch{}, extracts all sections, and applies them to the current AS.

The patch binary in Executable and Linking Format (ELF) lists changed sections between the unpatched and patched object files of the application to patch (for details about the patch-structure, we refer to the ``create-diff-object'' utility in \cite{kpatchGithub}).

\begin{figure}[tb]
\centering
\includegraphics[width=0.9\linewidth]{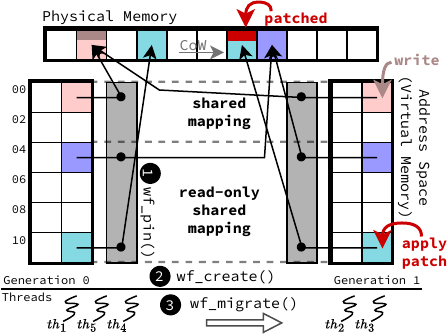}
  \caption{Process memory layout after \wfpatch{} operations.}
  \label{fig:asgen}
\end{figure}

\noindent\textbf{\wfpatch{} Operations.}
\Cref{fig:asgen} illustrates a process having two AS generations (initial AS on the left; cloned AS on the right) and five threads. Each box in memory represents a page, whereas an arrow pointing to it represents the translation process of the page table (gray bar next to the AS). The steps and the required \wfpatch{} operations to achieve the illustrated state are shown by a circled number with the respective operation.

Before the AS in \Cref{fig:asgen} is cloned, the pages in the region of addresses 4--11 are defined as read-only shared mapping (\circled{1}~\texttt{wf\_pin()}), while all other pages remain as shared mapping. Subsequently, the AS is cloned (\circled{2}~\texttt{wf\_create()}). The patch is applied to the new AS and affects the page at address~11, which is in the area of the read-only shared mapping. This results in the physical page to be copied. The change is ultimately applied to the copied page. AS generation~0 still points to the old unpatched physical page. Next, two threads migrate to the new AS generation (\circled{3}~\texttt{wf\_migrate()}). To highlight the shared mapping, a write to the page at address~1 is reflected by both AS generations, as both still point to the same physical page. The workflow of \wfpatch{} is to (1)~clone an AS, (2)~apply a patch, and (3)~individually migrate threads to the patched process version.

\subsubsection{\textbf{Technical Limitations}}
\label{sec:wfpatchLimitations}

\kpatch{} can only generate patches for applications written in C. Furthermore, the granularity of patches is based on functions, i.e.\ old functions are replaced with new ones. Additionally, a patch can only be applied to inactive functions, i.e.\ functions that are not currently active on the stack frame.

\wfpatch{} can only patch the read-only regions .text (executable code) and .rodata (initialized static constants). Thus, global variables or the layout of data structures cannot be patched.
Despite these limitations, a large number of patches can be applied in practice.
Rommel et al.~\cite{DBLP:conf/osdi/RommelDFKBMSL20} show in an analysis of more than 100 software fixes for six applications that about 87\% of patches are text-only.

\section{Related Work}
\label{sec:relatedWork}
We begin by examining the related work on live patching in the operating system community. These contributions, to the best of our knowledge, remain unexplored in database systems research. 

\noindent\textbf{Live Patching Research in the OS Community.}
Various (prototypical) user-space live patching frameworks exist~\cite{libpulp, DBLP:conf/osdi/RommelDFKBMSL20, DBLP:conf/middleware/WeichbrodtHAAK21, DBLP:conf/icse/ChenYCZY07, DBLP:conf/usenix/MakrisB09, DBLP:conf/oopsla/HaydenSDHF12, DBLP:conf/icde/HaydenSHF11, DBLP:conf/pldi/NeamtiuHSO06}.
Several enforce some form of global quiescence~\cite{DBLP:conf/oopsla/HaydenSDHF12, DBLP:conf/icde/HaydenSHF11, libpulp}, while others allow patch application at arbitrary points in time, but with the overhead of halting all threads~\cite{DBLP:conf/usenix/MakrisB09, DBLP:conf/icse/ChenYCZY07}. 

To our knowledge, \wfpatch{}~\cite{DBLP:conf/osdi/RommelDFKBMSL20} is the only framework
for live patching multi-threaded user-space applications 
with a wait-free patch application approach via local quiescence. 
The idea of multiple AS generations of \wfpatch{} has found applications beyond live patching, including thread-specific security~\cite{DBLP:conf/lctrts/Rommel00OL23} and execution variations (e.g., with and without tracing)~\cite{DBLP:conf/usenix/Tollner0ORL23}.

Under the hood, \wfpatch{} and \kpatch{}~\cite{kpatch} employ trampolines, a common technique used in live patching~\cite{DBLP:conf/icse/ChenYCZY07, DBLP:conf/uss/AltekarBBS05, DBLP:conf/osdi/RommelDFKBMSL20, kgraft,DBLP:conf/eurosys/ArnoldK09} to redirect function calls to the patched code once the control flow reaches it.

\noindent\textbf{Live Patching DBMS in the OS Community.}
Database systems are part of experimental evaluations of several live patching approaches~\cite{DBLP:conf/usenix/MakrisB09, DBLP:conf/osdi/RommelDFKBMSL20, DBLP:conf/oopsla/HaydenSDHF12, DBLP:conf/middleware/WeichbrodtHAAK21}, but are commonly treated as generic user-space applications, the same as caches or web servers. The unique challenges associated with database management systems are not considered, the database client experience is largely ignored. This can also be observed in the experiments of Rommel et al.~\cite{DBLP:conf/osdi/RommelDFKBMSL20}: They evaluated the \wfpatch{} framework based on six different multi-threaded user-space applications, including MariaDB (the only relational DBMS in their work). In the following, we discuss their experiments with MariaDB from database system perspective.

\emph{Request Latencies.}
Rommel et al.\ executed a customized benchmark against MariaDB with the one-thread-per-connection policy (thread connection policies are explained in \Cref{sec:mariaDBThreadModel}). Quiescence was triggered every 1.5~seconds, but without actually applying a patch. This experiment was performed for global and local quiescence, and the results for both quiescence methods are compared based on a histogram of measured client request latencies.

\emph{Runtime Penalty.}
Rommel et al.\ reported the runtime penalties of \wfpatch{} operations. They measured the overhead of AS cloning and AS switching for a single, fixed MariaDB configuration.

\emph{Discussion.}
Both experiments clearly show that the evaluation specifically focuses on \wfpatch{} and the concept of local quiescence (in comparison to global quiescence) rather than database systems: (1)~MariaDB is more complex and, in addition to the one-thread-per-connection policy, also supports a thread pool policy. (2)~Database systems are subject to different workload types such as OLTP or OLAP. (3)~The steps of loading and applying a real-world patch are not captured by the experiments. 
(4)~Different kinds of database systems (e.g.\ main memory vs.\ disk-based) have memory states of different size, and also differ in how they store data internally.

In summary, Rommel et al.\ designed their experiments to evaluate the benefits of \wfpatch{} and the local quiescence method exclusively from the perspective of operating systems research. Assuming the domain-specific perspective of database systems research, different experiments and analyses are required to assess live patching DBMSs along the desiderata outlined in \Cref{sec:challenges}. Given that their quiescence points for the thread pool connection policy are susceptible to deadlocks (see \Cref{sec:oltpWorkloads}), we propose a novel approach known as priority-based quiescence (detailed in \Cref{sec:implementationThreadpool}), which aims to ensure a safe and deadlock-free migration of threads within a thread pool.

\noindent\textbf{DBMS Upgrade Strategies.}
For applying a patch to a DBMS without the database clients noticing a downtime, a common approach is to perform a rolling upgrade, based on running additional instances~\cite{PostgreSQLBDR, PostgreSQLStandbyRolling, MicrosoftSQLServerRolling, AmazonUpgrade, DBLP:conf/sigmod/DagevilleCZAABC16, GaleraClusterRolling, MariaDBRolling}.
In this setup, the hardware costs multiply due to redundant provisioning of hardware and database instances. Furthermore, a rolling upgrade may, nevertheless, take time for a cluster to patch instance-by-instance. Furthermore, cluster performance is reduced during downtime, and the instance needs to recover its full memory state on startup.

\noindent\textbf{Live Patching DBMS in the Database Systems Community.}
There is little to no published research on live patching database systems. 
In a very early-work abstract, we conducted a first experiment on DBMS live patching~\cite{DBLP:conf/sigmod/Fruth22}. Since then, we have systematically extended our work to the point where we propose a novel solution for live patching with database thread pools. 

According to a blog entry~\cite{AzureLivePatching}, Azure SQL Database supports live patching since 2018. It employs an optimized C++ compiler for patch generation, and uses trampolines to redirect function calls to patched code. 
In particular, this approach is designed for Windows systems (whereas \wfpatch{} works for Linux). Furthermore, only few details about the solution for Azure SQL Database are known, and we found no published systematic experiments.

\noindent\textbf{DBMS Address Space Optimization.}
The novelty of \wfpatch{} comes from duplicating an address space. Its functionality is similar to Linux fork(), which is a common operator in today's database landscape to perform a snapshot of the memory state. Different database systems try to work around the overhead of fork() by, for example, reducing the number of page table entries by using larger page sizes~\cite{DBLP:conf/icde/KemperN11}. A recent proposal of asynchronous fork~\cite{DBLP:journals/pvldb/PangDBCSLXYWWLSYMG23} has been made to also reduce the fork() overhead of Redis~\cite{RedisFork}.

The exploitation of the address space or virtual memory has also found application in other areas, such as caching (e.g.\ DBMS buffer pool)~\cite{DBLP:journals/pacmmod/LeisA0L023} or query processing~\cite{DBLP:journals/pvldb/SchuhknechtDS16}.

\noindent\textbf{Priority Scheduling in DBMSs.}
Our novel contribution of priority-based quiescence designed for database thread pools, to be introduced in \Cref{sec:implementationThreadpool}, is independent of other priority mechanisms used within a database system, such as task scheduling~\cite{DBLP:journals/pvldb/PsaroudakisSMSA16, DBLP:conf/vldb/PsaroudakisSMA13}. In fact, our priority-based quiescence concept can seamlessly integrate with other mechanisms. For example, by aligning the priorities of quiescence with task priorities, it could be ensured that threads engaged in high-priority tasks are allowed to execute for at least as long as there are threads performing lower-priority tasks.

\begin{figure*}[t]
\centering
\includegraphics[width=\linewidth]{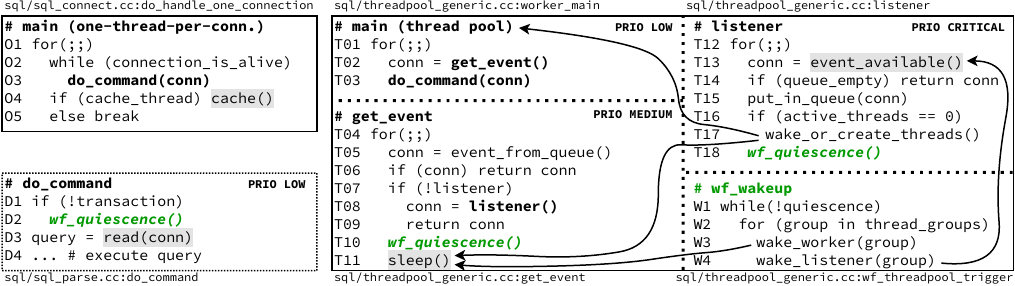}
  \caption{Implementing one-thread-per-connection (top left) and thread pool policy (right block), inspired by MariaDB.
  }
  \label{fig:codeOneThreadVsThreadpool}
\end{figure*}

\section{Safe Quiescence Points in Database Connection Management}
\label{sec:mariaDBThreadModel}

Database connection management is a carefully tuned component, and
developers need to take great care when injecting quiescence points. They must ensure that quiescence points are indeed safe states for threads to migrate.  
Ideally,
the quiescence points do not alter the connection management policy noticeably, and the same quiescence points can be used with
both global and local quiescence.

Next, we discuss the established connection management policies one-thread-per-connection and thread pool. 
We first consider quiescence points w.r.t.\ \emph{global quiescence}, which has the strictest requirements, and then discuss local quiescence. 

\subsection{One-Thread-per-Connection with Global Q.}
\label{sec:implementationOneThreadPerConnection}

\noindent\textbf{Preliminaries.} 
To introduce the one-thread-per-connection policy, we walk through a pseudocode implementation. While our pseudocode example is specifically inspired by the MariaDB source code\footnote{Based on git hash: \texttt{06fae75859}. The names and functions shown as pseudocode differ slightly from the original implementation, they were edited for easier readability.}, the process for setting quiescence points can be considered applicable in a more general context.

To the top left, \Cref{fig:codeOneThreadVsThreadpool} shows the main function of a worker thread. For now, we ignore the commands and functions set in italic green, which marks the injected code. 
Each connection is assigned a dedicated thread, its \emph{worker}. The worker executes commands, such as queries (line~\texttt{O3}), as long as the connection to the client is \emph{alive} (line~\texttt{O2}). Once the client connection is closed, the worker is either \emph{cached} (line~\texttt{O4}) or  terminated (line~\texttt{O5}). 
The \texttt{do\_command} function is the starting point for query processing: First, it performs a blocking read on the client connection (line~\texttt{D3}) and waits for input from the client. Once input is available, the query can be executed (line~\texttt{D4}). 

\noindent\textbf{Challenges.}
Quiescence points must avoid two threats:
(1)~\emph{Deadlocks:} A thread encountering a quiescence point will block and wait for the other threads to reach the barrier. If this thread already holds a lock to a data object, this can cause a deadlock when another thread requires this lock before it is able to reach its own quiescence point.
(2)~\emph{Starvation:} A thread that remains cached will not reach its quiescence point and therefore blocks all other threads that already wait at their barrier.

\noindent\textbf{Solution.}
The solution presented here is based on the approach proposed by Rommel et al.~\cite{DBLP:conf/osdi/RommelDFKBMSL20} (with minor refactoring) and addresses the following challenges:
(1) To not increase the risk of deadlocks, 
a worker thread must be outside of a transaction when it encounters a quiescence point.
Therefore, we check the transaction status (line~\texttt{D1} in \Cref{fig:codeOneThreadVsThreadpool}) before a quiescence point is reached (line~\texttt{D2}).
It is generally best practice to inject quiescence points high up in the call hierarchy, as only functions that are not currently active on the call stack can be patched at runtime. 
(2)~To prevent starvation, a patch request must cause all cached threads to wake up so that they may then reach their quiescence point. This wake-up call is triggered by the patcher thread (code not shown).

\noindent\textbf{Discussion.}
An inherent problem concerns blocking reads, where a thread waits for user input. In global quiescence, this will cause unbound wait times (line~\texttt{O3}), for example, with a client holding an idle connection. This problem can be addressed, for example, by interrupting the thread, as we also do with the listener in the thread-pool policy (discussed next).

\noindent\textbf{Implementation.} We adopted the solution based on Rommel et al.~\cite{DBLP:conf/osdi/RommelDFKBMSL20} and implemented it with minor changes in MariaDB.

\subsection{Thread Pool with Global Quiescence}
\label{sec:implementationThreadpool}

\noindent\textbf{Preliminaries.}
In the thread pool as implemented in MariaDB, thread groups are used to divide client connections into distinct sets. The size of the thread pool corresponds to the number of thread groups, whereby each thread group can consist of several threads with different roles (this is explained in more detail below). As the thread groups operate independently of each other, we assume that there is only one thread group in the following discussion.

We first assume that the database workload is high. Then, a dedicated listener thread manages a queue to distribute the work among the worker threads. Consequently, the listener adds connections to the queue when they have work available (lines \texttt{T13} and \texttt{T15} in \Cref{fig:codeOneThreadVsThreadpool}). 
Based on the producer-consumer model, worker threads dequeue connections from the queue (line \texttt{T05}) and then process the query (lines \texttt{T06} and \texttt{T03}).
The dedicated listener thread is only active in the \texttt{listener} function. With each iteration, it is checked whether an active thread can process the previously added event (line~\texttt{T16}). Otherwise, a sleeping thread is awakened or a new one is created (line~\texttt{T17};  see arrows in \Cref{fig:codeOneThreadVsThreadpool}). If the work queue is empty, the consuming worker thread goes to sleep (line~\texttt{T11}).

For a medium-to-low workload, there is no dedicated listener thread, but the worker threads temporarily assume this role:
A worker transitions to listener  (line~\texttt{T08}) and waits for a connection to become available to process data. Once a connection contains input, it is fetched (lines~\texttt{T14} and \texttt{T09}) and processed (line~\texttt{T03}). 

\noindent\textbf{Challenges.}
Since this policy is more complex, the risk of accidentally introducing deadlocks or allowing starvation is amplified.\footnote{Rommel et al.\ provide an implementation for the thread pool that deadlocks. This 
highlights the challenge in finding a functional solution.}
The quiescence points in which a thread pool has no dedicated listener is similar to the one-thread-per-connection policy. But specifically when MariaDB faces high loads and the thread pool has a dedicated listener,
we must carefully control the order in which threads may block upon reaching their 
quiescence points. 

Let us illustrate these risks and consider the scenario of a worker thread inside an active transaction waiting for the release of a lock.

\emph{Active Worker vs.\ Listener.}
A listener thread reaches its quiescence point and blocks. It no longer manages the queue. As long as the event which could release the desired lock is not added to the queue, the worker thread cannot complete its transaction. As quiescence points are purposefully positioned outside of transactions, the worker thread is indefinitely blocked.

\emph{Active Worker vs.\ Sleeping.}
A worker that awakes from sleep may have to be prevented from blocking when it encounters a quiescence point (line~\texttt{T10}). Otherwise, there may not be a worker left that handles events entering the queue, which could release the desired lock of the active worker.
This constitutes a deadlock.

\noindent\textbf{Novel Solution -- Priority-Based Quiescence.}
These scenarios motivate us to propose \emph{priority-based quiescence}, a priority-based approach to orchestrate blocking of threads. We assign priorities to threads depending on their current role, where threads with higher priority will only block at their quiescence point \emph{after} all lower-priority threads block. Put differently, a thread encountering a quiescence point skips it if there is still a lower-priority thread that has not yet reached its quiescence point. Intuitively, the higher-priority thread has not yet reached a state where blocking is \emph{safe}.

\emph{General Applicability.}
To adopt this approach, it is essential to identify the specific roles or tasks that a thread can perform. Subsequently, a hierarchy of roles needs to be established indicating the interdependencies. Prioritization is established on the basis of the following hierarchy: Tasks relying on others are assigned lower priorities, whereas tasks that are prerequisites for other tasks are given higher priorities (the hierarchy is reflected in the priorities). 
As a result, threads responsible for tasks on which other threads depend remain active until all these dependent threads are blocked. 

Following the assignment of priorities, careful consideration is required when placing quiescence points. Integration should guarantee that each thread consistently encounters quiescence points. Threads should pass a quiescence point only when not holding locks on shared resources. In scenarios with blocked or sleeping threads, common in a thread pool setup, an external mechanism (triggered by the \wfpatch{} thread) can awaken them, ensuring a reliable progression toward the quiescence point.

\emph{Adoption to MariaDB.}
In \Cref{fig:codeOneThreadVsThreadpool}, the priority of each role is annotated to the top right of each code block.  For MariaDB, three roles are identified with their respective priority: active worker (LOW), sleeping worker (MEDIUM) and listener (CRITICAL). The listener blocks last since it accepts incoming data upon which both workers depend. A sleeping worker is awakened to handle queries, supporting active workers in completing their tasks. Consequently, a sleeping worker should only block after active workers.

We have also injected a dedicated quiescence point for each role of a thread (line \texttt{T10} for a (sleeping) worker, line \texttt{T18} for a listener). To avoid problematic scenarios between blocked workers and sleeping workers, we trigger the \texttt{wf\_wakeup} method from the outside. It wakes all sleeping worker threads (line \texttt{W3}) and blocking listener (line \texttt{W4}) until global quiescence is reached (line \texttt{W1}).

\noindent\textbf{Discussion.}
The priority-based scheme is designed to prevent the deadlock scenarios outlined above.

\noindent\textbf{Implementation.}
We implemented our novel concept of priority-based quiescence for thread pools in MariaDB and extended the \wfpatch{} user-space library to support priorities.

\subsection{Adaption to Local Quiescence}
For both connection policies, the same quiescence points described above can also be utilized for local quiescence. We do not require adaptation since local quiescence is not plagued by the problems of global quiescence (bound/unbound wait times, deadlock). 
In fact, we could inject additional and \enquote{local quiescence specific} quiescence points in the source code. However, these benefits come with the limitation of reduced patchability, since local quiescence is limited to thread-local patches. Since global quiescence has the stricter requirements and local quiescence is compatible, we settle on a shared set of quiescence points in favor of less code complexity.

\section{Experiments}
\label{sec:experiments}

We evaluate live patching for database systems from the perspectives of the stakeholders discussed in Section~\ref{sec:challenges}.

\noindent\textbf{Hardware.}
Our server has two Intel Xeon Gold 6248R CPUs (24 cores per CPU; 3.0~GHz) and \SI{384}{\giga\byte} of main memory. To reduce system noise,
Intel Turbo-Boost is disabled. All CPU cores run at a fixed core frequency of 3.0~GHz. Since we can assign more than double the number of cores to each application (DBMS / benchmark framework) compared to the parallel queries being executed (detailed configuration given below), we have disabled Intel Hyper-Threading to utilize all 24 physical cores per CPU and to avoid competition for shared core cache.

\noindent\textbf{Live Patching Infrastructure.} The system runs Debian 11 with the latest \wfpatch{}\footnote{\label{footnote:mmview}\url{https://github.com/luhsra/linux-mmview}; git tag: \texttt{mmview-v5.15}}
Linux kernel (version 5.15). 
Live patching further requires the \wfpatch{} user-space library.

\noindent\textbf{Applications to be patched.} 
We extended the source code for the RDBMS MariaDB and the key-value store Redis. MariaDB implements the connection management policies of interest, and Redis allows us to easily control the size of the memory state.

\noindent\emph{\textbf{MariaDB.}}
\textbf{Code Extensions.}
We extended the MariaDB source code as discussed in Section~\ref{sec:mariaDBThreadModel}.
We injected quiescence points for the main thread, all worker threads, and dedicated listener threads. Therefore, we cover all threads relevant to transaction processing.\footnote{There are further threads in MariaDB which we do not live patch in our experiments. This is a simple technical limitation and can be easily resolved in productization.}

\noindent\textbf{Patches.}
We developed a fully automated pipeline to analyze the development history of an application for live-patchable code changes. For every version, we examine the modifications by checking if \kpatch{} can generate a patch and, upon success, proceed by automatically applying our source code changes, specifically injecting quiescence points.
This allows to obtain a wide range of patches and their various characteristics. Ideally, quiescence points would be integrated once and maintained consistently, while bug fixes would be implemented with live patching in mind~\cite{kpatchAuthorsGuide}. However, the technical limitations described in \Cref{sec:wfpatchLimitations} remain in effect.

MariaDB, written in C/C++, encounters limitations with the C-focused \kpatch{} tool for patch generation. Despite this, our automated pipeline, scanning versions 10.5.0--10.5.13 of MariaDB on GitHub, identified 117 live-patchable code changes. While all 117 patches contribute to our broader analysis (see \Cref{sec:patchApplicationOverhead}), we imposed two strong criteria for our in-depth evaluation: (1)~the patch must be officially labeled a ``bug'' in the MariaDB bug tracker, and (2)~it should modify a function in the stack trace below the \texttt{do\_command} function.  In consequence, the patch actually affects a function that is executed during a benchmark run (and not some dormant code, which is low risk to patch). The five selected patches are presented in \Cref{tab:mariadbPatches}. The first column provides a unique identifier used throughout this paper. In the PDF, the git hash and the corresponding Jira ticket of the code change are clickable links, pointing to the respective entries on GitHub and Jira, so that readers may inspect them in detail. The last two columns show the number of lines changed, excluding tests. These patches primarily resolve a bug with just a single line of code.

\begin{table}[tb]
  \caption{MariaDB patches fixing official bugs.}
\label{tab:mariadbPatches}
  
\begin{tabular}{lllrr}
ID & git hash & MariaDB Jira & \#LoC added & \#LoC deleted\\
\midrule
\#1 & \href{https://github.com/mariadb/server/commit/18502f99eb24f37d11e2431a89fd041cbdaea621}{18502f99eb} 
& \href{https://jira.mariadb.org/browse/MDEV-22185}{MDEV-22185} 
& +1 & -1\\
\#2  & \href{https://github.com/mariadb/server/commit/30d41c8102c36af7551b3ae77e48efbeb6d7ecea}{30d41c8102} 
& \href{https://jira.mariadb.org/browse/MDEV-22881}{MDEV-22881}
& +2 & -1\\
\#3 & \href{https://github.com/mariadb/server/commit/3bb5c6b0c21707ed04f93fb30c654caabba69f06}{3bb5c6b0c2} 
& \href{https://jira.mariadb.org/browse/MDEV-22113}{MDEV-22113} 
& +7 & -8\\
\#4 & \href{https://github.com/mariadb/server/commit/56402e84b5ba242214ff4d3c4a647413cbe60ff3}{56402e84b5}
& \href{https://jira.mariadb.org/browse/MDEV-21824}{MDEV-21824} 
& +1 & -1\\
\#5 & \href{https://github.com/mariadb/server/commit/5b678d9ea4aa3b5ed4c030a9bb5e7d15c3ff8804}{5b678d9ea4}
& \href{https://jira.mariadb.org/browse/MDEV-25251}{MDEV-25251} 
& +1 & -1\\
\end{tabular}
\end{table}

\noindent\textbf{Benchmarks.}
We adapted the benchmark harness BenchBase\footnote{\label{footnote:benchbase}\url{https://github.com/cmu-db/benchbase}; git hash: \texttt{979b53b043}} (formerly OLTP-Bench~\cite{DBLP:journals/pvldb/DifallahPCC13}) to trigger patch application. 
For OLTP workloads, we use the benchmarks NoOp, YCSB~\cite{DBLP:conf/cloud/CooperSTRS10} (scale factor 1,200) and TPC-C~\cite{TPCCSpecification} (scale factor ten). 
NoOp (No Operation) is extremely lightweight and just sends a single semicolon to the database.
For NoOp, we run BenchBase with the Epsilon Garbage Collector, as Java garbage collection can interfere with latency measurements~\cite{DBLP:conf/tpctc/FruthSMR21}.
For all OLTP benchmarks, a ten-second warm-up phase is followed by a 30-second benchmark phase.

For OLAP-style queries, we removed all OLTP queries from the benchmark CH-benCHmark~\cite{DBLP:conf/sigmod/ColeFGGKKKNNPSSSW11}. BenchBase is configured to use a 30~minute measurement phase without a warm-up phase, while triggering the patching process after five~minutes.

\noindent\textbf{Configurations.}
For all benchmarks against MariaDB, we use ten~terminals, i.e.\ ten parallel connections. MariaDB is executed with default settings, except for the thread pool which is limited to three thread groups. Connections are assigned round-robin to the three groups, each consisting of workers and potentially a dedicated listener.
We further reduce latency by disk I/O, placing the MariaDB data directory in a filesystem mapped to main memory.

We assign MariaDB to all 24~cores of CPU~1 and BenchBase to 23~cores of CPU~0. As ten connections are used, each thread processing queries can be scheduled on its own physical core, leaving more than 10~cores for background threads.

These configurations and (system) optimizations enable accurate and repeatable measurements.

\noindent\emph{\textbf{Redis.}}
\textbf{Code Extensions.}
For Redis (version 7.0.11), we injected one quiescence point in the single-threaded main event loop. 

\noindent\textbf{Patches.}
Redis is implemented in C, i.e.\ it is highly compatible for patch generation with \kpatch{}.
From the development history of Redis versions 5.0.0--7.0.11 on GitHub, we extracted 529 patches.

\noindent\textbf{Benchmarks.}
We extended the vendor benchmark framework redis-bench (part of the Redis project) to measure  
individual latencies. We use benchmarks consisting of only \texttt{SET} or \texttt{GET} operations.

\noindent\textbf{Expanded Charts.}  
\Cref{sec:appendix} includes charts for various experiments, displaying the complete data, with some charts enlarged for better clarity. Each figure caption in the following sections indicates whether an enlarged version of the chart and additional discussion of the results are available in the appendix.

\begin{figure}[tb]
\centering
\includeExperimentFigure{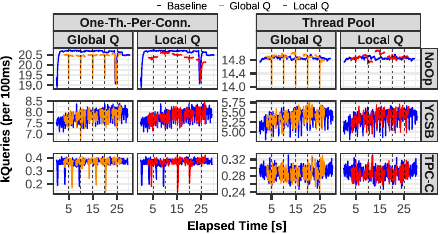}
  \caption{Query throughput over time for OLTP  workloads, comparing MariaDB without patch application (\enquote{baseline}, blue), patching in 5-second intervals with different setups.
  See \Cref{fig:appQps} (\Cref{sec:appOLTP}) for a chart displaying all five patch IDs. Bar charts showing the query throughput are also available in \Cref{fig:appBar} (\Cref{sec:appBarQps}).}
  \label{fig:qps}
\end{figure}

\subsection{Developer Perspective: Impact of Quiescence}
\label{sec:experimentsDevPerspective}

We assume the perspective of the database developer, concerned about the safety of quiescence points and performance regressions.

\subsubsection{\textbf{OLTP Workloads}}
\label{sec:oltpWorkloads}

Figure~\ref{fig:qps} shows throughput over time aggregated over bins of 100~ms, running OLTP-benchmarks against MariaDB. 
We compare one-thread-per-connection (left) against thread pool (right). We  patch the system (using patch ID~\#1) 5, 10, 15, 20, and 25 seconds into the benchmark (but each in a separate run) to catch the system in different states. We employ both global (left column; orange line) and local quiescence (right column; red line). 
We show the throughput over time 2~seconds before/after patch application (for a compact visualization and an easy comparison of the results). In each chart, the live patching run is superimposed onto the baseline run (blue line) for comparison.

The experiments for the remaining patches (patch IDs \#2--\#5) do not provide new insights. Additionally, the total throughput does not show any significant impact due to live patching. We refer to our artifacts for the corresponding charts.

\noindent\textbf{Results.}
Obviously, throughput is highest for the lightweight NoOp
benchmark and lower for more intense workloads. Throughput over time for live patching aligns closely with the baseline. Only a slight variation can be observed for the NoOp benchmark (e.g.\ about 100~queries per 100~ms difference). However, we consider these marginal differences to fall within the normal variations between individual runs.

Focusing on the NoOp benchmark, we can observe for all three configurations (baseline, global and local quiescence) of the one-thread-per-connection policy a short drop in throughput at about 25~seconds. However, for global quiescence, we can additionally observe a short drop in throughput for both connection policies when performing a live patch. These drops cannot be observed for local quiescence; therefore, they can be attributed to the blocking of threads for global quiescence. Unlike the NoOp benchmark, which is lightweight and sensitive, the YCSB and TPC-C benchmarks show fluctuating throughput over time, masking any temporary throughput decline due to global quiescence.

TPC-C has the lowest throughput for the one-thread-per-con\-nection policy with about 350~queries per 100~ms, thus it takes on average about 0.35~ms for one thread to process one query. In global quiescence, this is also the average time that a thread has to wait for all other threads to reach their quiescence point (after which the patch is applied). Thus, live patching under an OLTP workload does not noticeably impact throughput.

One concern from the developer perspective is that of encouraging deadlocks.
While we did do not encounter deadlocks in any run, this is not the case for the earlier implementation~\cite{WfPatchRepro} by Rommel et al.\ for the MariaDB thread pool policy, which does not employ the priority-based quiescence proposed by us. When we replicate our experiment using the YCSB and TPC-C benchmark with their implementation, every patch request inevitably causes a deadlock (in 100 out of 100 runs). This illustrates the intricacies of choosing the quiescence points and justifies our approach.

\subsubsection{\textbf{OLAP Workload}}
\label{sec:olapWorkload}
Global quiescence is unproblematic with short-lived queries, as the threads will frequently encounter their quiescence points. But with long-running queries, threads synchronizing at the quiescence barrier may incur longer wait times. 
Therefore, we discuss our experiment with our OLAP workload and the one-thread-per-connection policy.

\begin{figure}[tb]
\centering
\includeExperimentFigure{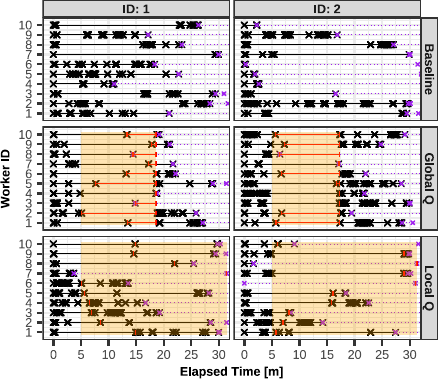}
  \caption{Fine-granular traces of the activities of 10 worker threads in MariaDB.
  Top row shows baseline. Yellow background frames the time from patch application until all threads have migrated to the patched version. See \Cref{fig:appChSingleLatencies} (\Cref{sec:appOlap}) for a chart displaying all five patch IDs.}
  \label{fig:chSingleLatencies}
\end{figure}

\noindent\textbf{One-thread-per-connection.}
\Cref{fig:chSingleLatencies} shows the fine-grained traces of ten worker threads in MariaDB running the one-thread-per-connection policy. 
The topmost row shows baseline runs (no patches applied). Randomness in issuing queries in the benchmark harness and differences in query runtimes
lead to unique traces.

Each line shows the activity of one given worker thread over time.
The start and end of a query are marked with a cross and a connecting line, respectively. A query that does not complete within the 30~minute measurement phase is shown as a small purple cross, connected by a dotted purple line. The pale yellow rectangle frames the synchronization time, that is, the time until all threads have migrated to the new version. Red bars connected by a red line show the time a thread blocks in a quiescence point until it is migrated to the patched version.

The second row shows traces of patch ID~\#1 and ID~\#2, applied with the global quiescence method.
The third row shows traces of patch ID~\#1 and ID~\#2 applied, but now with the local quiescence method.
Further traces are included in our artifacts.

\noindent\textbf{Results.}
We first focus on patch ID~\#1 under global quiescence (\Cref{fig:chSingleLatencies}, left middle chart): Worker~2 reaches its quiescence point about 5~seconds after the patch request is issued. It blocks for about 13~minutes until  worker~4 reaches its quiescence point (last thread to reach its barrier). Therefore, all workers that have reached their quiescence point before worker~4 have been idle and blocked. As expected, global quiescence incurs long wait times.

Local quiescence results in evidently higher concurrency (more crosses within the yellow-shaded area), as threads promptly migrate upon encountering a quiescence point. Yet, overall, it takes longer for all workers to complete their migration: As all worker threads keep working, they keep competing for resources (locks), and they can delay the other threads in reaching their quiescence points. In both bottom charts,
the last worker completes migration about one~minute \emph{after} the 30-minute benchmark window.

\subsubsection{\textbf{Synchronization Time}}
\label{sec:synchronizationTime}
To evaluate synchronization time for the thread pool policy, we perform an experiment using OLTP benchmarks and the source code version of patch ID~\#1 (patch ID \#2--\#5 show the same effect and the results are available in the artifacts). For the duration of the 30-second benchmark phase, we trigger patch application every 100~ms, i.e.\ 300 patch requests per run. We measure synchronization time, but \emph{without} applying a real patch. Patch application is highly patch-dependent (an effect which we explore below). For global quiescence, we measure the time until all threads reach their barrier and for local quiescence the duration of cloning the AS plus the time until all threads have migrated to the new AS generation. The experiment is conducted for a scale-out scenario, ranging the thread pool size from three to 20 (keep in mind that our benchmark utilizes ten parallel connections that are evenly distributed among all thread groups round-robin).

\begin{figure}[tb]
\centering
\includeExperimentFigure{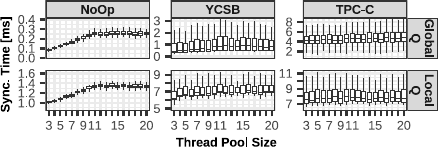}
  \caption{Boxplots of synchronization times, varying thread pool size and triggering patch application every 100~ms.
  See \Cref{fig:appThreadpoolPatchEvery} (\Cref{sec:appSyncTime}) for a chart displaying all five patch IDs and the one-thread-per-connection policy.
  }

  \label{fig:threadpoolPatchEvery}
\end{figure}

\begin{figure*}[t]
\centering
\includeExperimentFigure{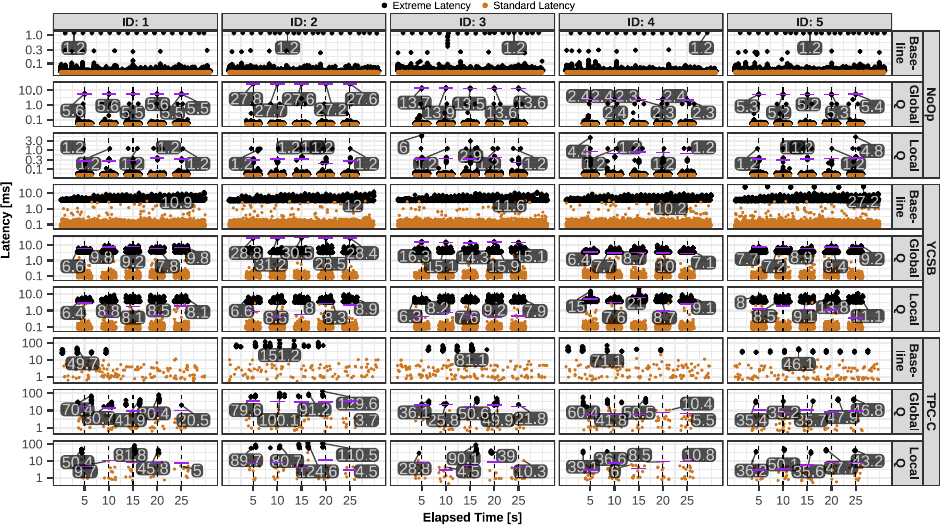}
  \caption{Latencies in live patching MariaDB under the one-thread-per-connection policy. Varying patches and setup.
  See \Cref{fig:appSingleLatenciesOneThreadPerConnection} (\Cref{sec:appLatencies}) for an enlarged version of this chart and \Cref{fig:appSingleLatenciesThreadPool} (\Cref{sec:appLatencies}) for the thread pool policy.}

  \label{fig:singleLatenciesOneThreadPerConnection}
\end{figure*}

The boxplots in \Cref{fig:threadpoolPatchEvery} show synchronization times, where outliers (data points outside the boundary of the whiskers; whiskers are based on the 1.5 IQR value) are not shown for better visualization. Columns specify the benchmark, while rows specify the quiescence method. The x-axis denotes  thread pool size, and the y-axis the synchronization time in milliseconds (y-axes scaled individually).

As can be expected, the synchronization time is inversely correlated with throughput (see \Cref{fig:qps}): The higher the throughput, the lower the synchronization time since quiescence points are passed more frequently.
However, this correlation is not the case for the OLAP workload visualized in \Cref{fig:chSingleLatencies}: The synchronization time for global quiescence is lower compared to local quiescence, even though throughput is lower. This advantage comes at the cost of blocking threads, leading to less competition for locks, but also to a reduced degree of parallelism.

For the NoOp benchmark in \Cref{fig:threadpoolPatchEvery}, synchronization time rises as the number of thread groups increases, but this is noticeable only up to a thread pool size of ten.  
This indicates that synchronization time increases with the number of thread groups, but this effect is significant only when the thread groups are active (i.e., have an assigned connection). Inactive thread groups do not deteriorate the synchronization time. A similar trend is observed for global quiescence in the YCSB and TPC-C benchmarks, albeit to a lesser extent due to the generally higher synchronization time.

For this experiment, no deadlock appeared for any of the approximately 81,000 patch applications with global quiescence. This once again highlights -- albeit empirically -- the applicability of our priority-based quiescence approach for thread pools.

\subsection{Client Perspective: Extreme Latencies}
\label{sec:experimentExtremeLatencies}

The experience of the database client is shaped by latencies in query processing, as
tail latencies can accumulate in distributed systems~\cite{DBLP:journals/cacm/DeanB13}.
Figure~\ref{fig:singleLatenciesOneThreadPerConnection} visualizes\footnote{For this sequence of charts, we adapted the visualization style and the scripts from a reproduction package for an earlier  project~\cite{DBLP:conf/tpctc/FruthSMR21}.}  query latencies in live patching MariaDB with the one-thread-per-connection policy. 
We compare five patches, different OLTP benchmarks, and quiescence methods.

\noindent\textbf{Baseline.}
The top row shows a baseline run for the NoOp benchmark where no patches are applied. Along the horizontal axis, we show progress over time. The dots represent latencies measured within BenchBase and reflect the experience of the database client. Latencies beyond the $99.95^\text{th}$ percentile are colored black (extreme values). The maximum latency is labeled. The standard latencies in orange are heavily sampled (down to 10\%) to reduce overplotting. 

Comparing the baseline runs (rows 1, 4 and~7), we confirm that the more intensive the workload, the higher the extreme latencies in the baseline runs. These charts not only underscore the robustness and repeatability of our experiments but also highlight the influence of code versions on performance. When examining charts for different patch IDs, a highly consistent latency pattern is apparent. However, occasional variations in extreme latencies arise (e.g., TPC-C baseline row) due to each MariaDB patch being associated with a specific code version, with an individual performance profile.

\noindent\textbf{Live Patching.}
We focus on the 2nd and 3rd charts (counting from the top) in the left column, showing live patching of MariaDB under the NoOp benchmark for patch ID~\#1.
We perform five isolated runs and issue a patch request at either 5, 10, 15, 20, or 25 seconds into the benchmark (to hit the system in different states).
The charts show the time slices 1.5~seconds before/after patch application (for a compact visualization and an easy comparison of the results).
By comparing the latencies at the time of patch application against the baseline run, we can observe the overhead of live patching.
Moreover, the horizontal purple lines visualize the elapsed time for all threads to successfully migrate.

\noindent\textbf{Results.}
We again focus on patch~\#1 and the NoOp benchmark. In the baseline run, we observe standard latencies of about 0.03~ms and extreme latencies in the 1~ms range. With global quiescence, we observe one extreme latency of about 5.6~ms at patch application time. This corresponds to the time it takes for all threads to migrate (purple line) and can clearly be attributed to patch application. Note that these extreme latencies include the time until all threads have reached their barrier and the time required to load and apply the patch (to be explored in \Cref{sec:patchApplicationOverhead}).
For local quiescence, we only observe a slight increase in tail latencies. The purple line shows the maximum time a thread needs to migrate, which is at about 0.3~ms.

Let us also compare the behavior of different patches.
The baseline runs are highly similar under the NoOp benchmark,
yet
different patches cause different extreme latencies. 
With global quiescence, the tail latencies of patch~\#2 are at about 27.5~ms and for patch~\#3 at about 13.5~ms. This suggests that the time for loading and applying the patch is patch-specific. We explore this effect in \Cref{sec:patchApplicationOverhead}. A similar behavior can be observed for YCSB, but not for TPC-C: This benchmark is more work-intensive, and the latency overhead is still within the standard range of TPC-C.

\subsection{Provider Perspective: Predictable Overheads}
\label{sec:patchApplicationOverhead}

For the database provider, it is crucial to be able to assess the factors influencing the overhead of live patching.

\noindent\textbf{Size of Database State.}
Rommel et al.~\cite{DBLP:conf/osdi/RommelDFKBMSL20} carefully explored key impact factors from the operating systems perspective. They measured the overhead for address space cloning and switching (both operations in local quiescence) within a tight loop across six different user-space applications, each executed with a fixed configuration. Their experiment shows that creating an address space scales with its size, while migration is a constant-time operation.

For the database provider, the former is a concern, since DBMSs commonly have large memory states. To explore this further, we turn to Redis, a memory-based key-value store where we can easily control memory consumption. For example, we can issue \texttt{SET} operations with a data size of \SI{400}{\kibi\byte}. The page size on our machine is \SI{4096}{\byte}, so each \texttt{SET} results in the allocation of 100~pages (neglecting other internal data structures). As the page table stores one pointer for each page (the page table entry), and a pointer has \SI{8}{\byte}, this results in \SI{800}{\byte} of new page table entries (PTEs). 
\texttt{SET} operations thus inflate the page table and serve as benchmark workload.

\begin{figure}[tb]
  \centering
  \includeExperimentFigure{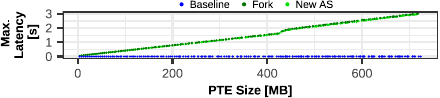}
  \caption{Impact of size of database state on  max.\ query latencies, comparing address space cloning and forking in Redis.}
  \label{fig:redis}
\end{figure}

We create 200~instances of Redis with a total page table size of up to \SI{818}{\mebi\byte}. Page table size can be used as a proxy metric for the number of page table entries (neglecting details of the internal tree structure), and correlates to main memory usage.
\Cref{fig:redis} visualizes the \emph{maximum} query latencies in each baseline run (blue ticks). These are not affected by the size of the memory state.
The (modified) patcher thread triggers Redis to clone an address space while Redis is under load. We capture the maximum query latency within a window of $\pm1$~second  (light green ticks). For reference, in a third experiment, the patcher thread triggers Redis to fork() instead of creating a new address space (dark green ticks). The results for address cloning and forking are near-indistinguishable. 

\noindent\textbf{Results.}
The maximum latencies increase linearly with the size of the page table, i.e.\ the allocated main memory. An administrator who knows the size of the allocated memory can predict the maximum latencies to be expected in live patching. However, even the delay in the range of seconds still outperforms a database restart, as it takes several minutes to restore \SI{120}{\giga\byte} of data from shared-memory~\cite{DBLP:conf/sigmod/GoelCGMMHW14}. It is also important to note that cloning an address space only needs to copy \SI{8}{\byte} (PTE) per \SI{4}{\kibi\byte} of data (page).

 AS cloning is implemented similarly to the fork() system call \cite{DBLP:conf/osdi/RommelDFKBMSL20}, which results in highly similar latencies. AS cloning also suffers from the process freeze of fork()~\cite{DBLP:conf/icde/KemperN11, RedisFork}: \emph{All} threads are stopped for copying the memory map structure, i.e.\ \emph{all} threads experience approx.\ the given latency (even if we only plot the maximum).

\noindent\textbf{Size of the Binary Patch.}
We measured the impact of the patch size on the patch application time for two systems.
For MariaDB, we use a total of 117~patches, for Redis a total of 529 patches, all real-world patches extracted from GitHub.
We only measure the patch application time from within the \wfpatch{} user-space library.

\begin{figure}[tb]
\centering
\includeExperimentFigure{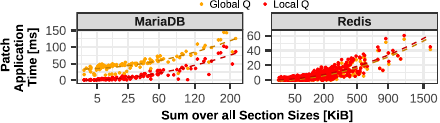}
  \caption{Impact of patch size on patch application time.}
  \label{fig:elf}
\end{figure}

\Cref{fig:elf} plots the size of the binary patches (stating the sum of their section sizes) against the time it takes to actually apply the patch. We compare the global and local quiescence method. Recall that in the first case, the patch is applied directly to the address space (in place); in the second case, the patch is applied to a cloned address space. 
In this experiment, barrier wait times etc.\ are ignored.
We put both databases under load (NoOp for MariaDB and \texttt{GET} for Redis), so that the threads reach their quiescence points.

\Cref{fig:elf} shows the relationship between the total sum of the size of all sections of the patch file (x-axis; please note square root scaling) and the patch application time, for global quiescence (orange) and local quiescence (red). A dashed regression line has been imposed for both quiescence methods. We observe that the time for patch application increases with the size of the patch. However, the latencies of MariaDB (one-thread-per-connection) are higher than those for Redis. For MariaDB, applying a patch with a total section size of about \SI{200}{\kibi\byte} takes about 50~ms using local quiescence and 85~ms using global quiescence. For Redis, latencies are in the range of 7~ms. This reveals that the patch application for MariaDB with local quiescence is faster than with global. This observation cannot be made for Redis.

\noindent\textbf{Results.} The duration of patch application depends on the size of the patch binary files. Moreover, for the multi-threaded MariaDB, patch application takes longer and the duration depends on the quiescence method used, which is not the case for Redis.

\section{Discussion and Outlook}
\label{sec:discussion}

Our study is the first to systematically evaluate the potential of live patching from the perspective of \emph{database systems research}. 
In the following, we summarize the insights and point out challenges.

\noindent\textbf{Main Insights.}
Our experiments show that live patching can indeed be a viable alternative to traditional means of updating, given that the patch is suitable. In our experiments, the quiescence points injected in the code for connection management did not cause any  deadlocks.\footnote{\label{revision:footnoteDeadlock}To \emph{guarantee} the absence of deadlocks, specialized methods such as code analysis, model checking, etc., must be employed, which is beyond the scope of this paper.}
This makes our priority-based quiescence in thread pools one of our core contributions.

Our experiments show that the observed tail latencies are in the milliseconds or lower second range. In comparison, a DBMS restart (1)~loses all connections, (2)~cannot respond or create new connections during the downtime and (3)~takes several minutes when restoring the database state from shared-memory, or even hours when restoring from disk (for \SI{120}{\giga\byte} \cite{DBLP:conf/sigmod/GoelCGMMHW14}).

{
\begin{table}[tb]
  \caption{Minimum and maximum latencies of live patching operations observed throughout the experiments.}
\label{tab:latencyBreakdown}
    \begin{tabular}{rrrr}
        & \multicolumn{1}{l}{Reach Quiescence} & \multicolumn{1}{l}{Apply Patch} & \multicolumn{1}{l}{New AS}\\
        \midrule
        Min. / Max. & \SI{0.4}{\micro\second} / \SI{26}{\minute} & \SI{0.1}{\milli\second} / \SI{145}{\milli\second} & \SI{1}{\milli\second} / \SI{3}{\second}
    \end{tabular}

\end{table}
}

\noindent\textbf{Latency Breakdown.}
Our experiments investigate the times taken by each individual live patching operation and their impacting factors.  
\Cref{tab:latencyBreakdown} shows the lowest and highest latencies for live patching operations observed in our experiments.
The overall latency of global quiescence comprises the maximum latency among individual threads reaching their quiescence point (``Reach Quiescence'' column), varying depending on the workload (\Cref{fig:chSingleLatencies} and \Cref{fig:threadpoolPatchEvery}), plus the subsequent loading and applying of the patch (``Apply Patch'' column), which is specific to the patch (\Cref{fig:elf}).
In case of local quiescence, the total latency comprises (1)~creating a new address space (``New AS'' column), which is proportional to the size of the memory state (\Cref{fig:redis}). (2)~The \wfpatch{} thread switches to the new address space, and (3)~applies the patch, both actions are executed in the background. (4)~Eventually, threads reach their quiescence point, and (5) switch to the patched address space. Switching address spaces is a constant-time operation~\cite{DBLP:conf/osdi/RommelDFKBMSL20}, in our experiments in the range of \SI{4}{\micro\second} -- \SI{2}{\milli\second}. 
In summary, the workload, the size of the DBMS memory state, and the specific patch independently influence latencies.

\noindent\textbf{Exploring Trade-Offs.}
For thread-local patches, we may choose between
the local or the global quiescence method.
Our experiments show that there is no direct answer to the question of which method to prefer. 
Regardless of the workload, global quiescence displays lower synchronization times. This can be favorable in terms of quickly closing software vulnerabilities. However, blocking threads reduce the degree of concurrency until global quiescence is reached. This negatively affects OLAP workloads in particular. In addition, there is a delay in loading and applying a patch, which depends on the patch size and also partly on the quiescence method. The duration of patch application is important not only for synchronization time, but also induces short-term tail latencies. After all, tail latencies can be noticeable, especially with OLTP workloads.

Main memory databases are particularly affected, especially by AS cloning of \wfpatch{} (required for local quiescence). The larger the state, the longer the entire process is frozen, which not only affects tail latencies, but also increases  synchronization time.

\noindent\textbf{Challenges for Database DevOps.}
Live patching requires custom Linux kernels and libraries. Such disruptive changes to the system stack necessitate careful testing.
Live patching is further disruptive to the database maintenance life cycle, and even affects the way code is changed. 
The Linux community already observes best practices on how to write code changes~\cite{kpatchAuthorsGuide} so that patches may be generated seamlessly.
Given a code change, DBMS vendors must carefully analyze any behavioral changes. More research is needed, employing program analysis to enable informed decisions.

\noindent\textbf{Challenges in Tooling.}
From our own experience, debugging is a serious challenge due to the acute lack of developer tools. Debuggers such as the GNU debugger (GDB) cannot be used out-of-the-box for applications having multiple ASs. If the application crashes, the core dump cannot be analyzed using Linux on-board tools. Over time, we can expect the tooling ecosystem to evolve with live patching in user-space applications becoming more common.

\noindent\textbf{Challenges in Patching Database Clusters.} So far, we have evaluated live patching a DBMS running as a single instance. Yet when a DBMS is deployed as multi-node installation, we might need new categories of patches (e.g., all threads of all instances must be in a safe state). Furthermore, cross-instance quiescence points may be required to synchronize nodes, for example, by deploying distributed synchronization algorithms such as two-phase commit. Further (technical) challenges may arise as a node must be able to process network packets until it is patched (to synchronize/communicate with other nodes). Utilizing priority-based quiescence could also be considered in this scenario. In addition, it may be necessary to implement priority-based quiescence at the cluster level to prevent cluster-wide deadlocks. This is particularly relevant in scenarios where nodes assume various roles, such as workers or coordinators, similar to the roles found within the thread pool policy.
A further question to be explored is how to  distribute patches in larger clusters, so that patches may be rolled out effectively.

Overall, having opened the field of research on live patching of databases, and having conducted our experiments and analyses, we see strong potential for follow-up research and real-world impact.

\bibliographystyle{ACM-Reference-Format}
\bibliography{bib}

\appendix
\section{Appendix Overview}
\label{sec:appendix}

This appendix serves as a supplementary resource to the experiments presented in \Cref{sec:experiments}. It provides enlarged plots and it includes all conducted experiments in full detail. 

Sections and figures are labeled with the prefix [new] or [extension to XY] to indicate whether they introduce a completely new chart or provide an extended version of an existing chart with additional details. 

\subsection{Addition to \Cref{sec:experimentsDevPerspective} (\nameref{sec:experimentsDevPerspective})}

\subsubsection{\textbf{[new] Average Throughput.}}
\label{sec:appBarQps}

\begin{figure*}[tb]
    \centering
    \includeExperimentFigure{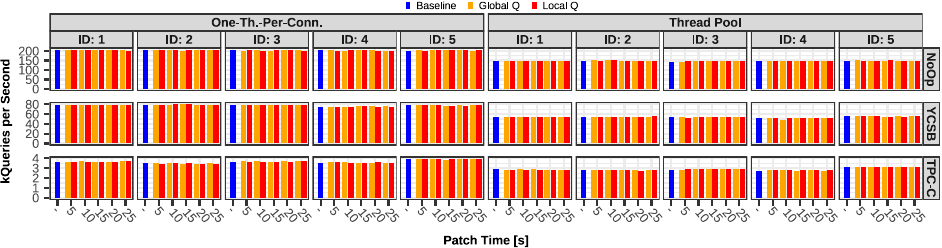}
\caption{[new] Average query throughput for OLTP workloads comparing MariaDB in three scenarios: without patch application (\enquote{baseline}, blue bars), with patch application using global quiescence (orange bars), and with local quiescence (red bars), across different patch application times.}
    \label{fig:appBar}
 \end{figure*}

\Cref{fig:appBar} presents bar charts showing the average throughput in 1,000 queries per second (QPS) for various OLTP workload scenarios. The figure compares the one-thread-per-connection (left) and the thread pool (right) policy. Patches are applied at 5, 10, 15, 20, and 25 seconds into the benchmark, each in a separate run, to capture the system in different states. This sequence is repeated for all five patches (columns). Each patch is applied once with global quiescence (orange bar) and once with local quiescence (red bar). The blue bar represents the baseline run with no patch applied.

\noindent\textbf{Results.}
Throughput is highest for the lightweight NoOp benchmark and lower for more intense workloads. Generally, the one-thread-per-connection policy results in higher throughput compared to the thread pool policy, with minimal variations between different MariaDB versions (i.e., different patch IDs). For example, the QPS for the TPC-C benchmark ranges from approximately 3,500 QPS (column ``ID: 2'') to about 3,900 QPS (column ``ID: 5''). These differences in throughput are observed only between different MariaDB versions and not within the same version (i.e., for the baseline and various live patching runs). Therefore, these differences can be attributed to variations between MariaDB versions.

The throughput for live patching closely aligns with the baseline performance. The minor variations observed between the baseline and live patching runs fall within the expected range of individual run variations. For instance, in the NoOp benchmark with patch ID 1, we observe a variation of approximately 1,000 QPS, which is negligible given the throughput of about 200,000 QPS.

In summary, live patching has no noticeable impact on the average query throughput.

\subsubsection{\textbf{[extension] Addition to \Cref{sec:oltpWorkloads} (\nameref{sec:oltpWorkloads})}}
\label{sec:appOLTP}

\begin{figure*}[tb]
\centering
\includeExperimentFigure{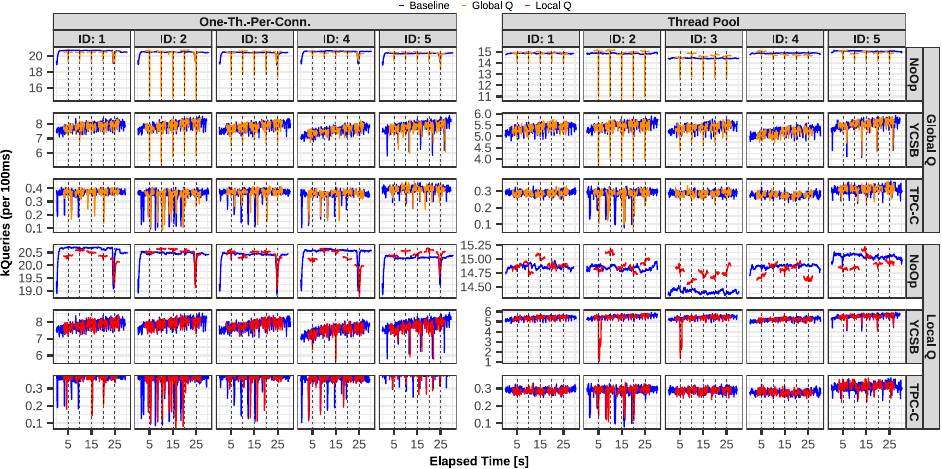}
  \caption{[extension to \Cref{fig:qps}] Query throughput over time for OLTP workloads, comparing MariaDB without patch application (\enquote{baseline}, blue), patching in 5-second intervals with different setups.}
  \label{fig:appQps}
\end{figure*}

\Cref{fig:appQps} shows an extension to \Cref{fig:qps}. The extended plot displays the throughput over time, aggregated into bins of 100~ms for all five patches (columns). For a detailed description of the visual elements of the chart, please refer to \Cref{sec:oltpWorkloads}.

\noindent\textbf{Results.}
Focusing on the NoOp benchmark, varying drops in throughput can be observed across different patches. For instance, the minimum throughput during the five patching runs for patch ID~\#1 is approximately 19,000 queries per 100~ms, while for patch ID~\#2, it drops to about 14,500 queries per 100~ms. A similar pattern is evident with the thread pool policy and is also observed for the YCSB benchmark. For the TPC-C benchmark, this effect is less noticeable as it shows a greater natural variance (e.g., see blue line for patch ID~\#2).

With local quiescence, no throughput drops attributable to live patching can be observed; any existing drops in throughput are due to inherent MariaDB performance variations, as these occur even in the baseline run (e.g., for the NoOp benchmark and the one-thread-per-connection policy at about 25~seconds into the benchmark).

\subsubsection{\textbf{[extension] Addition to \Cref{sec:olapWorkload} (\nameref{sec:olapWorkload})}}
\label{sec:appOlap}

\begin{figure*}[tb]
\centering
\includeExperimentFigure{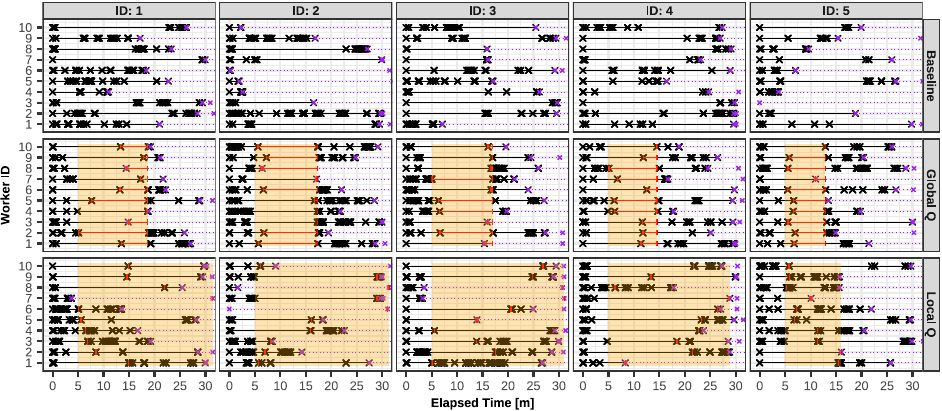}
  \caption{[extension to \Cref{fig:chSingleLatencies}] Fine-granular traces of the activities of 10 worker threads in MariaDB.
  Top row shows baseline. Yellow background frames the time from patch application until all threads have migrated to the patched version.}
  \label{fig:appChSingleLatencies}
\end{figure*}

\Cref{fig:appChSingleLatencies} shows an extension to \Cref{fig:chSingleLatencies}. The extended plot displays fine-grained traces of ten worker threads in MariaDB running the one-thread-per-connection policy for all five patches (columns). For a detailed description of the visual elements of the chart, please refer to \Cref{sec:olapWorkload}.

\noindent\textbf{Results.}
For all five patches, it can be observed for global quiescence that all threads complete their migration to the patched version at approximately 17~minutes into the benchmark. During global quiescence, each thread blocks at its quiescence point, waiting for all other threads to reach theirs. As fewer threads compete for resources, queries are processed more quickly, allowing quiescence points to be reached more timely. This ultimately accelerates the overall patching process.

In contrast, with local quiescence, the migration of threads progresses until the end of the benchmark (as seen in patch ID~\#1 to patch ID~\#4), but this approach allows queries to be processed in parallel the entire time, achieving a higher overall throughput.

\subsubsection{\textbf{[extension] Addition to \Cref{sec:synchronizationTime} (\nameref{sec:synchronizationTime})}}
\label{sec:appSyncTime}

\begin{figure*}[tb]
\centering
\includeExperimentFigure{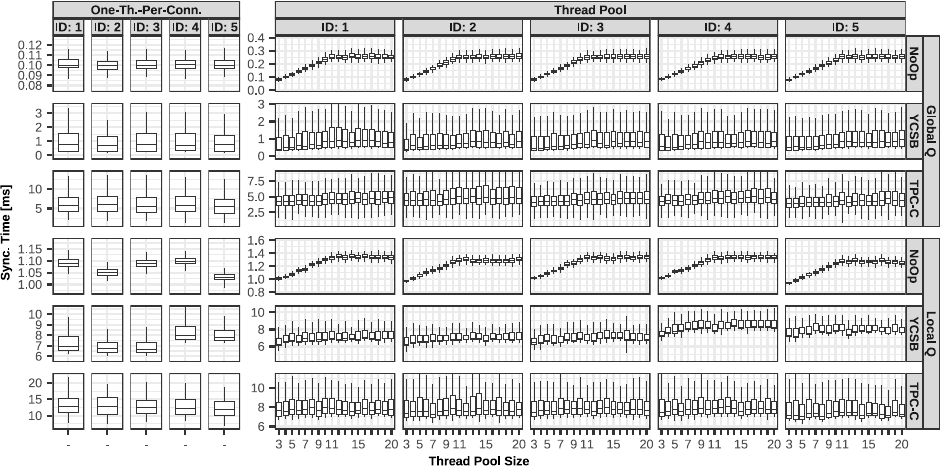}
  \caption{[extension to \Cref{fig:threadpoolPatchEvery}] Boxplots of synchronization times for the one-thread-per-connection policy (left) and varying thread pool sizes (right). Triggering patch application every 100~ms.}

  \label{fig:appThreadpoolPatchEvery}
\end{figure*}

\Cref{fig:appThreadpoolPatchEvery} presents an extension to \Cref{fig:threadpoolPatchEvery}, displaying boxplots of synchronization times for both the one-thread-per-connection policy and varying thread pool sizes for the thread pool policy. Boxplots for all five patches are included. For a detailed description of the visual elements of the chart, please refer to \Cref{sec:synchronizationTime}.

\noindent\textbf{Results.}
For the thread pool policy (right in \Cref{fig:appThreadpoolPatchEvery}), the different patches do not provide new insights as they show the same effect: Synchronization time increases with the size of active thread groups. This is particularly evident for the NoOp benchmark.

For completeness, synchronization time is also shown for the one-thread-per-connection policy (left in \Cref{fig:appThreadpoolPatchEvery}), where thread group scaling is not applicable (as no thread groups exist). Aside from minor variances between individual MariaDB versions (i.e. patch IDs), no notable effects are observed.

\subsubsection{\textbf{[extension] Addition to \Cref{sec:experimentExtremeLatencies} (\nameref{sec:experimentExtremeLatencies})}}
\label{sec:appLatencies}

In the following, we present the results of individual client request latencies, with particular focus on extreme latencies, for the one-thread-per-connection policy and the thread pool policy.

\begin{figure*}[tb]
\centering
\includeExperimentFigure{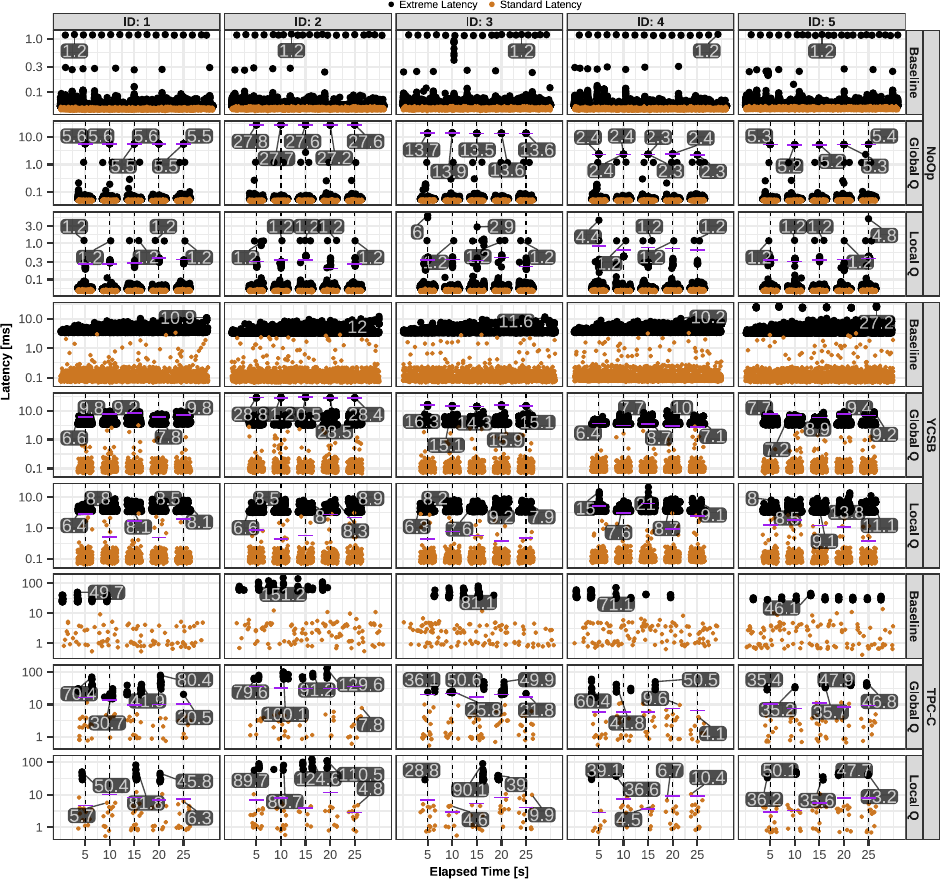}
  \caption{[extension of \Cref{fig:singleLatenciesOneThreadPerConnection}] Latencies in live patching MariaDB under the \emph{one-thread-per-connection} policy. Varying patches and setup.}

  \label{fig:appSingleLatenciesOneThreadPerConnection}
\end{figure*}

\paragraph{\textbf{One-thread-per-connection.}}
\Cref{fig:appSingleLatenciesOneThreadPerConnection} shows an enlarged version of \Cref{fig:singleLatenciesOneThreadPerConnection}, with all elements displayed in a larger format for better visibility. Therefore, we refer to \Cref{sec:experimentExtremeLatencies} for a detailed description of the visual elements of the chart and the discussion of the results.

\begin{figure*}[tb]
\centering
\includeExperimentFigure{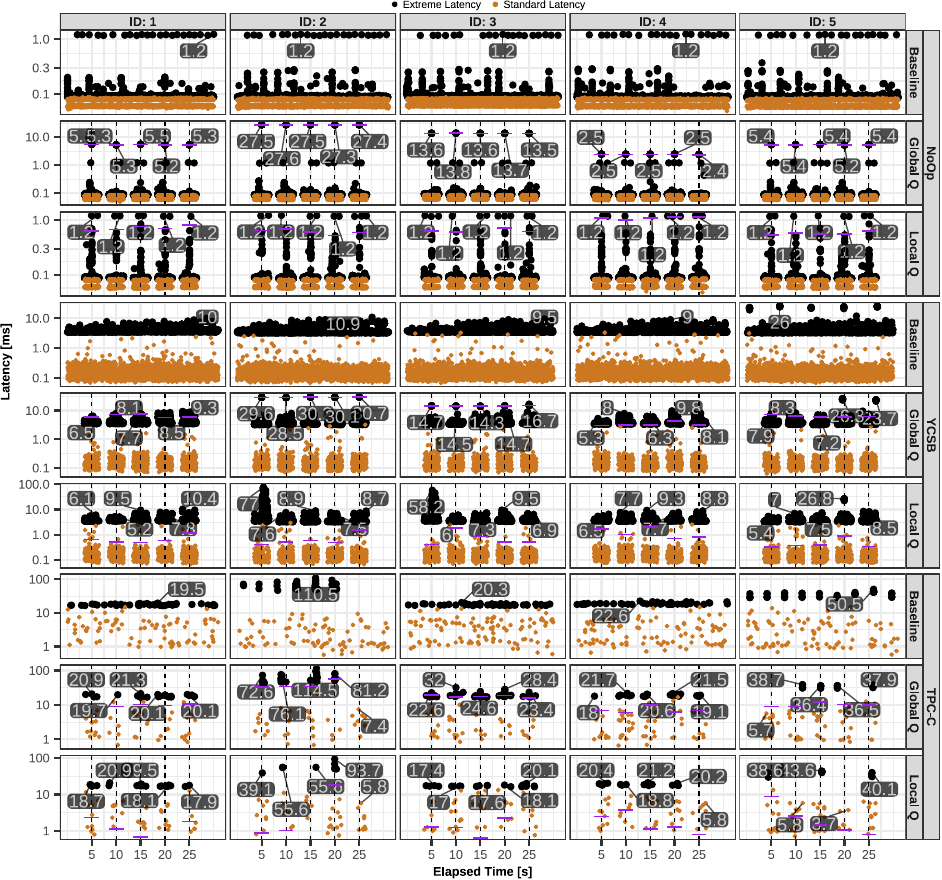}
  \caption{[new] Latencies in live patching MariaDB under the \emph{thread pool} policy. Varying patches and setup.}

  \label{fig:appSingleLatenciesThreadPool}
\end{figure*}

\paragraph{\textbf{Thread Pool.}}
\Cref{fig:appSingleLatenciesThreadPool} shows single request latencies for the thread pool policy. For a detailed description of the visual elements of the chart, please refer to \Cref{sec:experimentExtremeLatencies}.

\noindent\textbf{Results.}
Similar to the one-thread-per-connection policy (\Cref{sec:experimentExtremeLatencies}, \Cref{fig:appSingleLatenciesOneThreadPerConnection}), the baseline run for the NoOp benchmark shows maximum latencies around 1.2~ms. Under global quiescence and for the NoOp benchmark, each patch introduces a distinct maximum latency, such as 5.3~ms for patch ID~\#1 and 27.5~ms for patch ID~\#2. These latencies account for the time required for all threads to reach their quiescence point, as well as the time needed to load and apply the patch.
The YCSB benchmark also shows latency spikes caused by live patching, though these are not noticeable in all patch versions. Only patch ID~\#2 (around 30ms) and patch ID~\#3 (around 14ms) exhibit noticeable latency increases, surpassing the baseline maximum latency of 10ms.
In contrast, the TPC-C benchmark does not exhibit noticeable latency spikes using global quiescence because its higher baseline latency masks latencies of this magnitude. For example, the 27.5~ms maximum latency from patch ID~\#2 is not substantial compared to the baseline latency range of up to 110~ms.

With local quiescence, an increase in extreme latencies can be observed during patch application for the NoOp benchmark, along with a rise in the maximum latency. Since this increase occurs at request latencies around 0.3~ms, it remains unnoticeable in the YCSB and TPC-C benchmarks.

In summary, the extreme latency behavior of the thread pool policy mirrors that of the one-thread-per-connection policy, with noticeable impacts primarily during patch application in the benchmarks that have lower base latencies to start with.

\end{document}